\documentclass[aps,pra,showpacs,twocolumn]{revtex4-1}
\usepackage{amsmath}
\usepackage{amssymb}
\usepackage{graphicx}
\usepackage{epstopdf}
\usepackage{times,txfonts}
\usepackage{easybmat}
\usepackage{mathtools}
\newcommand{\ket}[1]{|#1\rangle}

\usepackage[bookmarks=false]{hyperref}
\hypersetup{colorlinks=true,citecolor=blue,linkcolor=blue,urlcolor=blue,pdfstartview=FitH,bookmarksopen=true}
\usepackage{color,soul}

\begin{document}

\title{Nonadiabatic holonomic quantum computation with Rydberg superatoms}
\author{P. Z.  Zhao}
\affiliation{Department of Physics, Shandong University, Jinan 250100, China}
\author{X. Wu}
\affiliation{Department of Physics, Shandong University, Jinan 250100, China}
\author{T. H. Xing}
\affiliation{Department of Physics, Shandong University, Jinan 250100, China}
\author{G. F. Xu}
\email{sduxgf@163.com}
\author{D. M. Tong}
\email{tdm@sdu.edu.cn}
\affiliation{Department of Physics, Shandong University, Jinan 250100, China}
\date{\today}
\pacs{03.67.Lx, 03.67.Pp, 03.65.Vf}

\begin{abstract}
Nonadiabatic holonomic quantum computation has received increasing attention due to its robustness against control errors as well as high-speed realization. Several schemes of its implementation have been put forward based on various physical systems, each of which has some particular merits. In this paper, we put forward an alternative scheme of nonadiabatic holonomic quantum computation, in which a universal set of quantum gates is realized based on Rydberg superatoms. A Rydberg superatom is a mesoscopic atomic ensemble that allows for only a single Rydberg excitation shared by many atoms within a blockade radius and can be used to generate the collective states to encode the qubits. In our scheme, the qubit is encoded into two collective ground states of Rydberg superatoms and the interaction between two long-range Rydberg superatoms is mediated by a microwave cavity with the aid of two additional collective Rydberg states. Different from the previous schemes, which are based on the systems in the microscope scale, the present scheme is based on atomic ensembles in the mesoscopic scale. Besides the common merits of nonadiabatic holonomic quantum computation such as the robustness and the speediness, the Rydberg-superatom-based scheme has the following particular merits: the long coherence time of Rydberg states and the operability of the mesoscopic systems.
\end{abstract}

\maketitle

\section{Introduction}
Nonadiabatic holonomic quantum computation \cite{Sjoqvist2012,Xu2012} is based on nonadiabatic non-Abelian geometric phases \cite{Anandan}, which exist in the quantum system that possesses a subset of states satisfying both the cyclic evolution and parallel transport conditions. Since  nonadiabatic non-Abelian geometric phases are only dependent on evolution paths and independent of evolution details, nonadiabatic holonomic gates are robust against control errors. Besides this, nonadiabatic non-Abelian geometric phases need not require the long run-time evolution that is necessary for adiabatic non-Abelian geometric phases \cite{Wilczek}, and therefore nonadiabatic holonomic gates allow high-speed realization, which is different from the adiabatic holonomic quantum computation \cite{Zanardi,Duan}. Due to the merits of both robustness against control errors and high-speed realization, nonadiabatic holonomic quantum computation has received increasing attention
\cite{Johansson2012,Spiegelberg2013,Zhang2014,Mousolou2014,Xu2015,Sjovist2016,Sjovist2016PRA,Sun2016,
Liang2014,Zhou2015,Xue2015,You2016,Xue2016,Xue2017,Xue2017PRA,Zhao,Zhao2017,Xu2017,Xu2017PRA,Mousolou2017,Xia2018,
Long,Abdumalikov,Arroyo,Duan2014,Zhou2017,Long2017}.
Several schemes of its implementation have been proposed based on various physical systems \cite{Liang2014,Zhou2015,Xue2015,You2016,Xue2016,Xue2017,Xue2017PRA,Zhao,Xia2018}, and each of them has some particular merits. Nonadiabatic holonomic quantum computation has been experimentally demonstrated with nuclear magnetic resonance \cite{Long,Long2017}, superconducting circuits \cite{Abdumalikov}, and nitrogen-vacancy centers in diamond \cite{Arroyo,Duan2014,Zhou2017}.

The practical implementation of nonadiabatic holonomic quantum computation requires coherent manipulation of a large number of coupled quantum systems. Thus finding quantum systems easy to be manipulated is particularly necessary. Considering the manipulation of the object in the mesoscopic scale is easy; here we examine the possibility of devising a Rydberg-superatom-based scheme of nonadiabatic holonomic quantum computation. A Rydberg superatom is a mesoscopic atomic ensemble that allows for only a single Rydberg excitation shared by many atoms within a blockade radius \cite{Lukin2001}. It is simpler to prepare mesoscopic Rydberg superatoms with an array of traps than to prepare a single atom in each trap \cite{Tong2004,Heidemann2007,Bakr2009,Ebert2015}. The stable collective ground states of the Rydberg superatom can be taken as the well-defined qubit states, and the encoded qubit is more robust against the atom leakage than the qubit encoded by the single atom states \cite{Dur2000}. In addition, the long lifetime of the high-lying collective Rydberg states facilitates the manipulation of quantum systems within coherence time. Due to the above attractive features, we think that Rydberg superatoms are a competitive candidate for implementing nonadiabatic holonomic quantum computation. In fact, due to the above attractive features, Rydberg superatoms have been widely used for many other quantum information processing tasks \cite{Brion2007,Muller2009,Han2010,Wu2010,Dudin2012,Beterov2013,Weber2015,Zeiher2015,Sarkany2015}. In this paper, we aim to realize nonadiabatic holonomic quantum computation by using Rydberg superatoms.

It is worth noting that Rydberg atoms have already been used to implement nonadiabatic geometric quantum computation \cite{Zhao} as well as nonadiabatic holonomic quantum computation \cite{Xia2018}. Different from these previous schemes, which  are implemented by using Rydberg blockade regime \cite{Jaksch2000} with single atoms, the present scheme of nonadiabatic holonomic quantum computation is to use microwave-cavity-mediated interaction with Rydberg superatoms. As stated above, Rydberg superatoms are with a mesoscopic scale, having the merit of operability. However, a mesoscopic scale may lead to the weak dipole-dipole interaction between the nonadjacent Rydberg superatoms when the number of superatoms is large, and it further spoils the Rydberg blockade regime for the long-range Rydberg superatoms. Thus, if a Rydberg-superatom-based scheme can be effectively performed by using the Rydberg blockade regime, the number of Rydberg superatoms needs to be limited in a small scale, which hinders the scalability of quantum computation because scalable quantum computation requires coherent manipulation of a large number of coupled quantum systems.  To avoid this problem, we use a microwave cavity with the aid of two additional collective Rydberg states to mediate the interaction between Rydberg superatoms.  This causes our scheme to not only allow for the manipulation of the object in a mesoscopic scale, but also to avoid the limitation on the number of Rydberg superatoms.

In our scheme, we encode the qubit by a pair of collective ground states of Rydberg superatoms and then realize a universal set of nonadiabatic holonomic gates acting on the collective ground states. The one-qubit gates are performed by using off-resonant laser pulses. The nontrivial two-qubit gate is realized with the aid of a microwave cavity, where the transitions between the two collective ground states and two collective Rydberg states are facilitated by exchanging virtual photons through the common cavity mode.
The paper is organized as follows. In Sec. II, we demonstrate how to prepare the needed collective states. In Sec. III, we realize two noncommuting one-qubit nonadiabatic holonomic gates based on Rydberg superatoms. In Sec. IV, we realize a nontrivial two-qubit nonadiabatic holonomic gate based on Rydberg superatoms. In Sec. V, we discuss the feasibility of our scheme. Section V is the conclusion.

Before proceeding further, we briefly explain how a holonomic gate can be obtained. Consider a $M$-dimensional quantum system defined by Hamiltonian $H(t)$.  Its unitary operator can be expressed as $U(t)=\mathrm{T}\exp{[-i\int_0^tH(t')dt']}$, where $\mathrm{T}$ is time ordering. Assume there exists a $L-$dimensional subspace $\mathcal{S}(t)=\mathrm{Span}\{|\psi_{k}(t)\rangle\}^{L}_{k=1}$ with $|\psi_{k}(t)\rangle=U(t)|\psi_{k}(0)\rangle$, and $\mathcal{S}(0)$ is taken as the computational space. Then, the unitary transformation $U(\tau)$ is a nonadiabatic holonomic gate acting on the
$L-$dimension subspace if $\ket{\psi_k(t)}$ fulfill the following requirements  \cite{Sjoqvist2012,Xu2012}:
\begin{align}
&(\mathrm{i})~ \sum^{L}_{k=1}|\psi_{k}(\tau)\rangle\langle\psi_{k}(\tau)|=
\sum^{L}_{k=1}|\psi_{k}(0)\rangle\langle\psi_{k}(0)|, \notag\\
&(\mathrm{ii})~~\langle\psi_{k}(t)|H(t)|\psi_{m}(t)\rangle=0, ~k,m=1,2,\cdot\cdot\cdot,L.
\end{align}
Condition $(\mathrm{i})$ guarantees that the evolution of the subspace is cyclic, while $(\mathrm{ii})$ ensures that $U(\tau)$ is purely geometric on the subspace.

\section{Preparation of the collective states}

To perform a nonadiabatic holonomic gate, one needs a space with at least three dimensions, where a two-dimensional subspace is used as the computational space while the other dimensions are auxiliaries.
To construct a three-dimensional space, we consider a Rydberg superatom consisting of $N$ identical four-level atoms, each of which has three stable ground states $|g\rangle$, $|0\rangle$, and $|1\rangle$, and a Rydberg state $|r\rangle$. Here, the ground state $|g\rangle$  acts as an initialized state that can be used to generate the needed collective states. As shown in Ref. \cite{Lukin2001}, all the atoms can be trapped in the ground state $|g\rangle$, and thus the Rydberg superatom can be initially prepared in a collective ground state $|\bar{g}\rangle=|g_{1}\cdot\cdot\cdot g_{N}\rangle$, where $|g_{k}\rangle$ represents the ground state $|g\rangle$ of the $k$th Rydberg atom. We first drive all Rydberg atoms from the ground state $|g\rangle$ to the Rydberg state $|r\rangle$. Due to Rydberg blockade, only a single Rydberg atom can be excited from the ground state to the Rydberg state, and as a result, one can realize the collective Rydberg state,
\begin{align}
|\bar{r}\rangle=\frac{1}{\sqrt{N}}\sum^{N}_{k=1}|g_{1}g_{2}\cdot\cdot\cdot r_{k}\cdot\cdot\cdot g_{N}\rangle,
\end{align}
where $|r_{k}\rangle$ represents the Rydberg state of the $k$th Rydberg atom.
We then drive the Rydberg atoms from the Rydberg state $|r\rangle$ to the ground states $|0\rangle$ or $|1\rangle$, resulting from which two collective ground states of the Rydberg superatom,
\begin{align}
|\bar{0}\rangle=\frac{1}{\sqrt{N}}\sum^{N}_{k=1}|g_{1}g_{2}\cdot\cdot\cdot 0_{k}\cdot\cdot\cdot g_{N}\rangle, \notag\\
|\bar{1}\rangle=\frac{1}{\sqrt{N}}\sum^{N}_{k=1}|g_{1}g_{2}\cdot\cdot\cdot 1_{k}\cdot\cdot\cdot g_{N}\rangle,
\end{align}
are realized. Here, $|0_{k}\rangle$ and $|1_{k}\rangle$ represent the ground states $|0\rangle$ and $|1\rangle$ of the $k$th Rydberg atom, respectively.
Thus, the needed collective states are prepared. A three-dimensional space, spanned by $\{|\bar{0}\rangle,|\bar{1}\rangle,|\bar{r}\rangle\}$, is constructed, where $\{|\bar{0}\rangle,|\bar{1}\rangle\}$ are used as the computational basis while $|\bar{r}\rangle$ acts as an auxiliary. It is worth noting that the collective ground state $|\bar{g}\rangle$ cannot be used as the computational basis because the excitation of $|\bar{g}\rangle$ can cause an undesired excitation of $|\bar{0}\rangle$ or $|\bar{1}\rangle$. This is the reason why we do not use the previous three-level setup of Rydberg superatoms \cite{Lukin2001,Ebert2015}, where the three-dimensional space consists of $|\bar{g}\rangle$, $|\bar{1}\rangle$, and $|\bar{r}\rangle$. In the following, we will realize a universal set of nonadiabatic holonomic gates based on above encoding.

\section{One-qubit gates}

\begin{figure}[t]
   \includegraphics[scale=0.3]{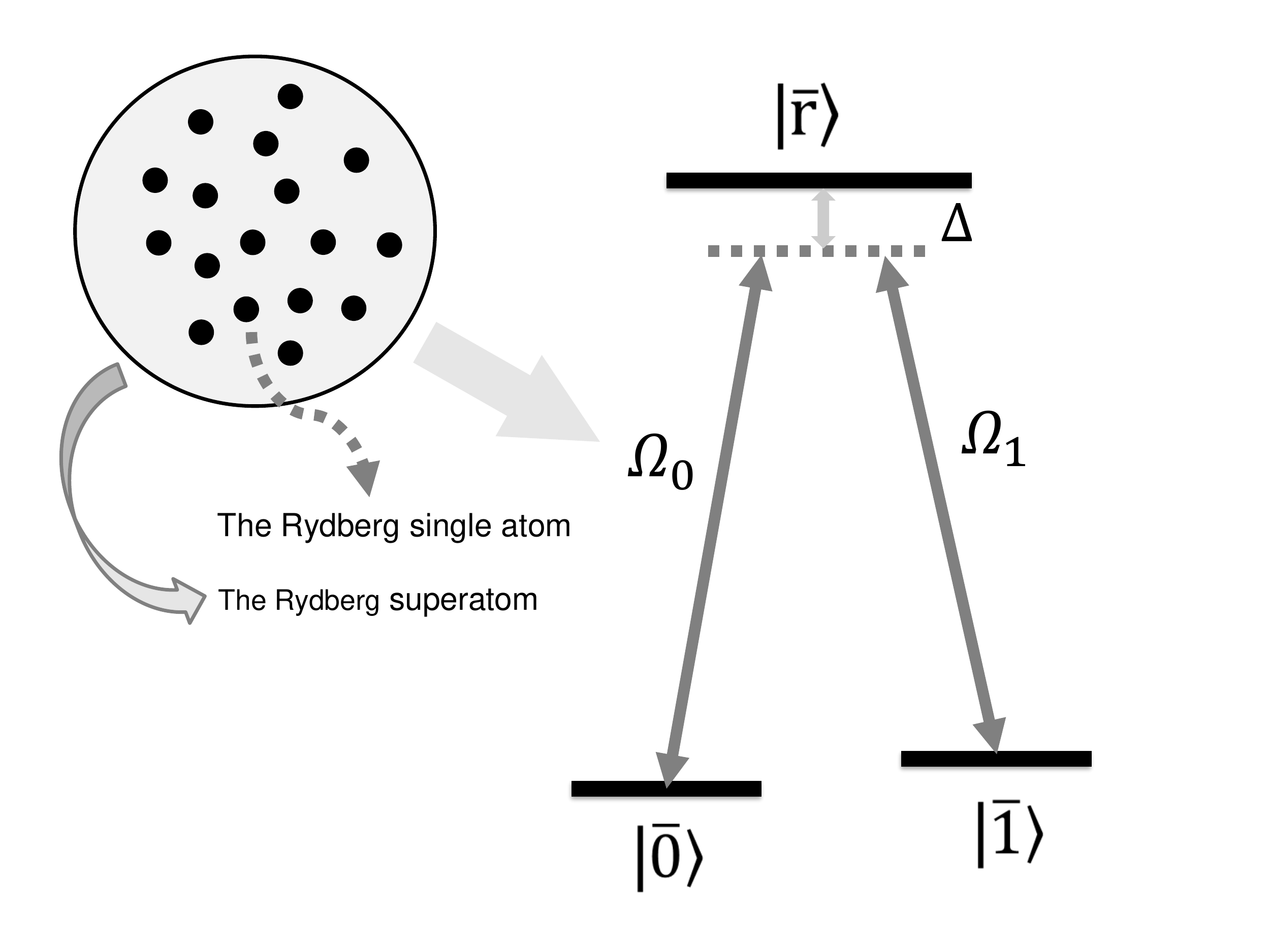}
   \caption{Setup for $\Lambda$ configuration of the Rydberg superatom. A Rydberg superatom consists of
    two collective ground states $|\bar{0}\rangle$ and $|\bar{1}\rangle$, and a collective Rydberg state $|\bar{r}\rangle$. The transitions $|\bar{0}\rangle\leftrightarrow|\bar{r}\rangle$ and $|\bar{1}\rangle\leftrightarrow|\bar{r}\rangle$ are facilitated by off-resonant laser pulses with the same detuning $\Delta$, but different Rabi frequencies $\Omega_{0}(t)$ and $\Omega_{1}(t)$.} \label{Fig1}
\end{figure}
We first demonstrate how to realize one-qubit nonadiabatic holonomic gates with a Rydberg superatom. As illustrated in Sec. II, the Rydberg superatom consists of two collective ground states $|\bar{0}\rangle$ and $|\bar{1}\rangle$ and a collective Rydberg state $|\bar{r}\rangle$. We drive the transitions $|\bar{0}\rangle\leftrightarrow|\bar{r}\rangle$ and $|\bar{1}\rangle\leftrightarrow|\bar{r}\rangle$ by off-resonant laser pulses with the same detuning $\Delta$ but different  Rabi frequencies $\Omega_{0}(t)$ and $\Omega_{1}(t)$, shown in Fig. \ref{Fig1}.
In the rotating frame, by using the rotating wave approximation, the Hamiltonian describing the Rydberg superatom interacting with the laser pulses reads
\begin{widetext}
\begin{align}
H(t)=\Delta|\bar{r}\rangle\langle\bar{r}|+\left[\Omega_{0}(t)|\bar{r}\rangle\langle\bar{0}|
+\Omega_{1}(t)|\bar{r}\rangle\langle\bar{1}|+\mathrm{H.c.}\right],
\end{align}
where $\mathrm{H.c.}$ represents the Hermitian conjugate terms.

To realize our nonadiabatic holonomic gates, we use square laser pulses. The common detuning and the two Rabi frequencies are set as
$\Delta=2\Omega\sin\theta$, $\Omega_{0}(t)=\Omega\cos\theta\cos(\varphi/2)
$, and $\Omega_{1}(t)=\Omega\cos\theta\sin(\varphi/2)$, where $\Omega$, $\theta$, and $\varphi$ are time-independent parameters \cite{Xu2015,Sjovist2016}. In this case, the Hamiltonian is time independent, and the evolution operator $U(t)=\exp(-iHt)$  can be written as, in the basis $\{|\bar{0}\rangle,|\bar{1}\rangle,|\bar{r}\rangle\}$,
\begin{align}
U(t)=\left(
  \begin{array}{ccc}
   \sin^{2}\frac{\varphi}{2}+(\cos\phi_{t}  +i\sin\phi_{t}\sin\theta)\cos^{2}\frac{\varphi}{2}e^{-i\phi_{t}\sin\theta} &
   \sin\frac{\varphi}{2}\cos\frac{\varphi}{2}
   [(\cos\phi_{t}+i\sin\phi_{t}\sin\theta)e^{-i\phi_{t}\sin\theta}-1]   &
   -i\sin\phi_{t}\cos\theta\cos\frac{\varphi}{2}e^{-i\phi_{t}\sin\theta}\\
   \sin\frac{\varphi}{2}\cos\frac{\varphi}{2}
   [(\cos\phi_{t}+i\sin\phi_{t}\sin\theta)e^{-i\phi_{t}\sin\theta}-1]  &
   \cos^{2}\frac{\varphi}{2}+(\cos\phi_{t}
   +i\sin\phi_{t}\sin\theta)\sin^{2}\frac{\varphi}{2}e^{-i\phi_{t}\sin\theta} &
   -i\sin\phi_{t}\cos\theta\sin\frac{\varphi}{2}e^{-i\phi_{t}\sin\theta}\\
   -i\sin\phi_{t}\cos\theta\cos\frac{\varphi}{2}e^{-i\phi_{t}\sin\theta} &
   -i\sin\phi_{t}\cos\theta\sin\frac{\varphi}{2}e^{-i\phi_{t}\sin\theta} &
   (\cos\phi_{t}-i\sin\phi_{t}\sin\theta)e^{-i\phi_{t}\sin\theta} \\
  \end{array}
\right),
\end{align}
where $\phi_{t}=\Omega t$.
If the evolution period $\tau$ satisfies
\begin{align}
\phi_{t}=\Omega\tau=\pi, \label{period}
\end{align}
the evolution operator $U(\tau)$ reads
\begin{align}
U(\tau)=\left(
  \begin{array}{ccc}
   \sin^{2}\frac{\varphi}{2}+\cos^{2}\frac{\varphi}{2}e^{-i\pi(1+\sin\theta)} & \sin\frac{\varphi}{2}\cos\frac{\varphi}{2}[e^{-i\pi(1+\sin\theta)}-1] & 0\\
   \sin\frac{\varphi}{2}\cos\frac{\varphi}{2}[e^{-i\pi(1+\sin\theta)}-1] & \cos^{2}\frac{\varphi}{2}+\sin^{2}\frac{\varphi}{2}e^{-i\pi(1+\sin\theta)} & 0\\
   0 & 0 & e^{-i\pi(1+\sin\theta)}\\
  \end{array}
\right).
\end{align}
\end{widetext}
The above equation shows that a state initially prepared in the computational space $\mathcal{S}=\mathrm{Span}\{|\bar{0}\rangle,|\bar{1}\rangle\}$ evolves back to $\mathcal{S}$ after the whole evolution, i.e., the condition $(\mathrm{i})$ is satisfied. With the aid of the commutation relation $[H,U(t)]=0$, one can also show that the parallel transport condition $(\mathrm{ii})$,
\begin{align}
\langle i(t)|H|l(t)\rangle=\langle i|H|l\rangle=0,~~|i\rangle,|l\rangle\in\mathcal{S},
\end{align}
is satisfied, where $|i(t)\rangle=U(t)|i\rangle$ and $|l(t)\rangle=U(t)|l\rangle$.
So, the evolution operator $U(\tau)$ is a nonadiabatic holonomic gate acting on the computational space $\mathcal{S}$.

In the following, we demonstrate that arbitrary one-qubit gates can be realized by using the evolution operator $U(\tau)$.
One can see that if the initial states are confined to the computational subspace $\mathcal{S}$, the evolution operator $U(\tau)$ is equivalent to
\begin{align}
U(\tau)=e^{-i\frac{\pi}{2}(1+\sin\theta)}e^{-i\frac{\pi}{2}(1+\sin\theta)\boldsymbol{n\cdot\sigma}},
\end{align}
where $\boldsymbol{n}=(\sin\varphi,\cos\varphi)$ is a unit vector and  $\boldsymbol{\sigma}=(\sigma_{x},\sigma_{z})$ is the standard Pauli operator acting on $|\bar{0}\rangle$ and $|\bar{1}\rangle$. By setting $\varphi=0$ and $\varphi=\pi/2$, we can obtain two noncommuting one-qubit rotation gates around the $Z$ and $X$ axes, respectively. Thus, arbitrary one-qubit nonadiabatic holonomic gates acting on the computational subspace $\mathcal{S}$ can be realized.

\section{The two-qubit gate}

To realize nonadiabatic holonomic quantum computation, besides one-qubit gates, a nontrivial two-qubit gate is needed. We now demonstrate how to realize a nontrivial two-qubit nonadiabatic holonomic gate with Rydberg superatoms.

\begin{figure}[t]
  \includegraphics[scale=0.4]{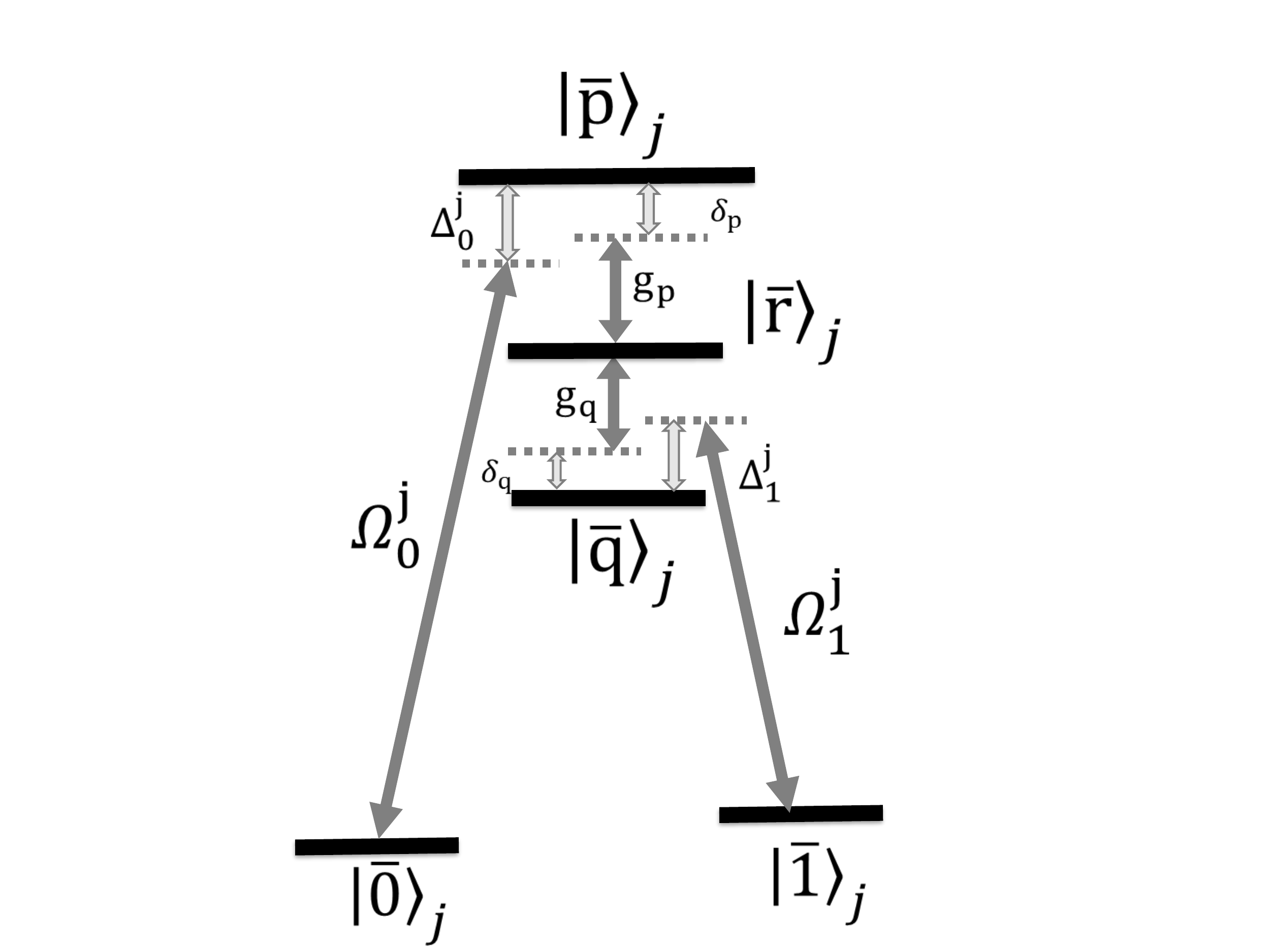}
  \caption{Configuration of the $j$th $(j=1,2)$ Rydberg superatom for the implementation of the two-qubit gate. The quantum system consists of two identical Rydberg superatoms, each of which has two collective ground states $|\bar{0}\rangle$ and $|\bar{1}\rangle$, and three adjacent collective Rydberg states $|\bar{p}\rangle$, $|\bar{r}\rangle$, and $|\bar{q}\rangle$. The transitions between the adjacent collective Rydberg states, $|\bar{p}\rangle\leftrightarrow|\bar{r}\rangle$
  ($|\bar{r}\rangle\leftrightarrow|\bar{q}\rangle$), are facilitated by a common cavity mode with coupling constants $g_{p}$ ($g_{q}$) and detunings $\delta_{p}$ ($\delta_{q}$). The transitions
  $|\bar{0}\rangle_{j}\leftrightarrow|\bar{p}\rangle_{j}$ ($|\bar{1}\rangle_{j}\leftrightarrow|\bar{q}\rangle_{j}$)
  of the $j$th individual Rydberg superatom, are driven by off-resonant laser pulses with Rabi frequencies $\Omega^{j}_{0}(t)$ [$\Omega^{j}_{1}(t)$] and detunings $\Delta^{j}_{0}$ ($-\Delta^{j}_{1}$).}\label{Fig2}
\end{figure}

\begin{figure}[t]
  \includegraphics[scale=0.35]{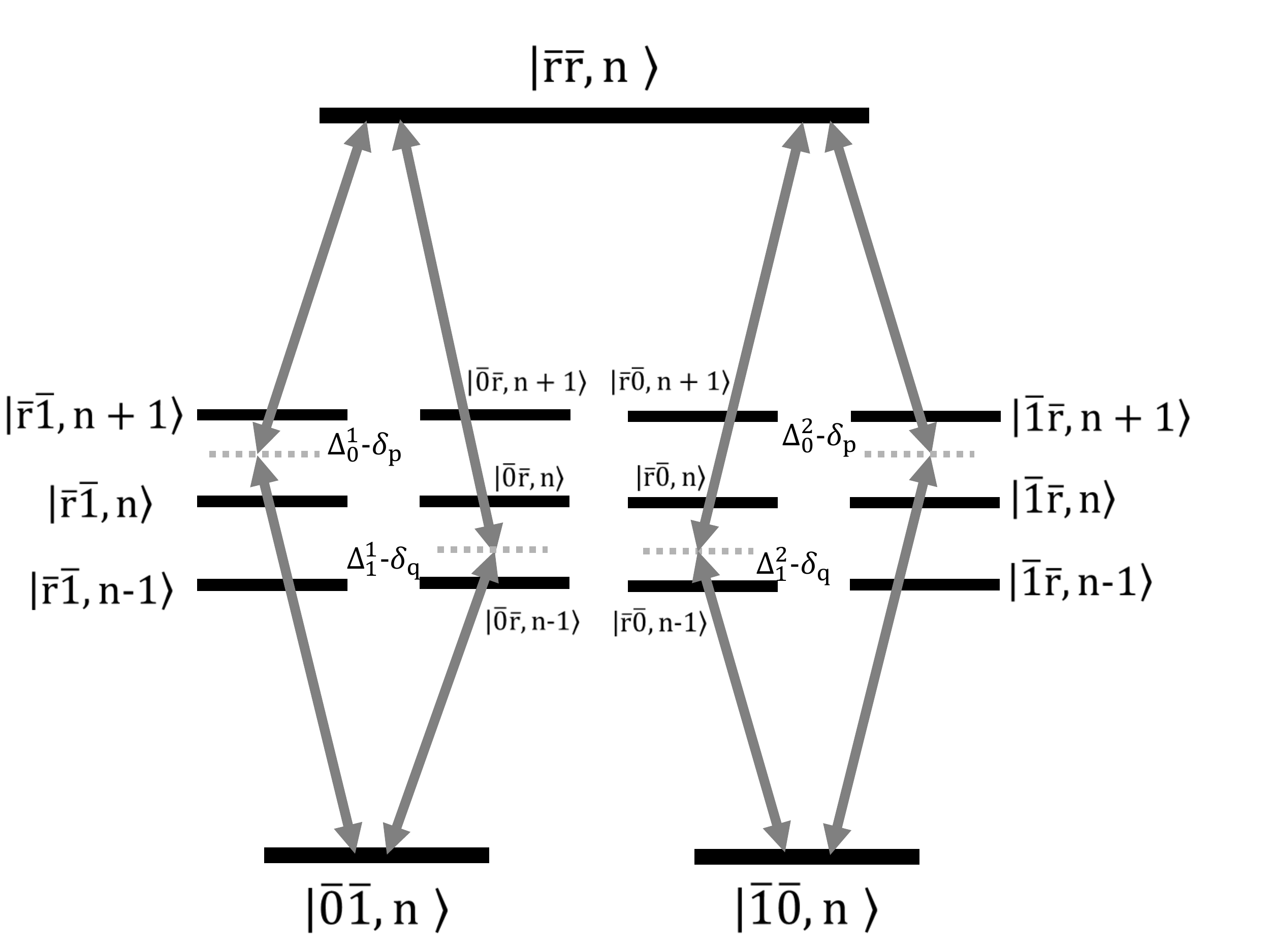}
  \caption{Setup for the effective coupling of a pair of Rydberg superatoms. When the collective Rydberg states $|\bar{p}\rangle$ and $|\bar{q}\rangle$ are decoupled and the conditions in Eqs. (\ref{condition}) and (\ref{condition1}) are satisfied, the single-atom transitions $|\bar{0}\rangle_{j}\leftrightarrow|\bar{r}\rangle_{j}$ and $|\bar{1}\rangle_{j}\leftrightarrow|\bar{r}\rangle_{j}$ can be strongly suppressed due to the large detuning conditions while the double-atom transitions $|\bar{0}\bar{1}\rangle\leftrightarrow|\bar{r}\bar{r}\rangle$ and $|\bar{1}\bar{0}\rangle\leftrightarrow|\bar{r}\bar{r}\rangle$ are allowed by exchanging virtual photons between the superatoms through a common cavity mode, where the photon number states $|n\rangle$, $|n-1\rangle$, and $|n+1\rangle$ have been used in the setup.} \label{Fig3}
\end{figure}

Consider a pair of Rydberg superatoms prepared in two spatially separated traps with the blockade interaction between the Rydberg superatoms being zero.
To perform nonadiabatic holonomic gates, we need to realize a three-level setup coupling of two Rydberg superatoms. For this, besides the three collective states $|\bar{0}\rangle$, $|\bar{1}\rangle$, and $|\bar{r}\rangle$ used in the one-qubit gates, we need to introduce two additional collective Rydberg states $|\bar{p}\rangle$ and $|\bar{q}\rangle$,
\begin{align}
|\bar{p}\rangle&=\frac{1}{\sqrt{N}}\sum^{N}_{k=1}|g_{1}g_{2}\cdot\cdot\cdot p_{k}\cdot\cdot\cdot g_{N}\rangle,\notag\\
|\bar{q}\rangle&=\frac{1}{\sqrt{N}}\sum^{N}_{k=1}|g_{1}g_{2}\cdot\cdot\cdot q_{k}\cdot\cdot\cdot g_{N}\rangle,
\end{align}
where $|p_{k}\rangle$ and $|q_{k}\rangle$ represent the Rydberg states $|p\rangle$ and $|q\rangle$ of the $k$th Rydberg atom, respectively. Here, $|\bar{p}\rangle$ and $|\bar{q}\rangle$ can as well play the role of $|\bar{r}\rangle$ for the one-qubit gates.
The configuration of the $j$th $(j=1,2)$ Rydberg superatom is show in Fig. \ref{Fig2}.
Here, $|\bar{p}\rangle$, $|\bar{r}\rangle$, and $|\bar{q}\rangle$ are three adjacent collective Rydberg states.
The transitions between the adjacent collective Rydberg states,
$|\bar{p}\rangle\leftrightarrow|\bar{r}\rangle$ (  $|\bar{r}\rangle\leftrightarrow|\bar{q}\rangle$ ), are facilitated by a common cavity mode with coupling constants $g_{p}$ ($g_{q}$) and detunings $\delta_{p}$ ($\delta_{q}$) \cite{Sarkany2015}. The transitions $|\bar{0}\rangle_{j}\leftrightarrow|\bar{p}\rangle_{j}$ ($|\bar{1}\rangle_{j}\leftrightarrow|\bar{q}\rangle_{j}$)
of the $j$th individual Rydberg superatom are driven by off-resonant laser pulses with Rabi frequencies $\Omega^{j}_{0}(t)$ [$\Omega^{j}_{1}(t)$] and detunings $\Delta^{j}_{0}$ ($-\Delta^{j}_{1}$).
By using the rotating frame and the rotating wave approximation, the Hamiltonian of the system  reads
\begin{align}
\mathcal{H}(t)=&\sum_{j=1,2}\Bigg\{\left[\Omega^{j}_{0}(t)e^{i\Delta^{j}_{0}t}
|\bar{p}\rangle_{jj}\langle\bar{0}|
+\Omega_{1}^{j}(t)e^{-i\Delta^{j}_{1}t}|\bar{q}\rangle_{jj}\langle\bar{1}|+\mathrm{H.c.}\right] \notag\\
&+\left(g_{p}e^{i\delta_{p}t}a|\bar{p}\rangle_{jj}\langle\bar{r}|
+g_{q}e^{-i\delta_{q}t}a^{\dagger}|\bar{q}\rangle_{jj}\langle\bar{r}|+\mathrm{H.c.}\right)\Bigg\}, \label{hamiltonian}
\end{align}
where $a$ and $a^{\dagger}$ are the annihilation and creation operators of the cavity mode, respectively. If the large detuning conditions
$\Delta^{j}_{0},\delta_{p}\gg |\Omega^{j}_{0}(t)|,g_{p}$ and
$\Delta^{j}_{1},\delta_{q}\gg |\Omega ^{j}_{1}(t)|,g_{q}$
are satisfied, and the difference between the detunings $\delta_{p}$ and $\delta_{q}$ is sufficiently large compared to the coupling constants $g_{p},g_{q}$, i.e., $|\delta_{p}-\delta_{q}|\gg g_{p},g_{q}$, then the single-atom transition $|\bar{p}\rangle\leftrightarrow|\bar{q}\rangle$ and the two-atom transitions $|\bar{p}\bar{q}\rangle\leftrightarrow|\bar{r}\bar{r}\rangle$ and $|\bar{q}\bar{p}\rangle\leftrightarrow|\bar{r}\bar{r}\rangle$ will be avoided, and the collective Rydberg states $|\bar{p}\rangle$ and $|\bar{q}\rangle$ will be decoupled from the computational space $\mathcal{S}^{\prime}=\mathrm{Span}\{|\bar{0}\bar{0}\rangle,|\bar{0}\bar{1}\rangle,
|\bar{1}\bar{0}\rangle,|\bar{1}\bar{1}\rangle\}$. We use the approach given in Ref. \cite{James2007} to reduce the Hamiltonian $\mathcal{H}(t)$ in Eq. (\ref{hamiltonian}). The reduced Hamiltonian can be written as
\begin{widetext}
\begin{align}
\mathcal{H}^{\prime}(t)=
&\sum_{j=1,2}\left[
-\frac{g_{p}\Omega^{j}_{0}(t)}{2}\left(\frac{1}{\Delta^{j}_{0}}
+\frac{1}{\delta_{p}}\right)
e^{i(\Delta^{j}_{0}-\delta_{p})t}
a^{\dagger}|\bar{r}\rangle_{jj}\langle\bar{0}|
+\frac{g_{q}\Omega^{j}_{1}(t)}{2}\left(\frac{1}{\Delta^{j}_{1}}
+\frac{1}{\delta_{q}}\right)
e^{-i(\Delta^{j}_{1}-\delta_{q})t}a|\bar{r}\rangle_{jj}\langle\bar{1}|
+\mathrm{H.c.}\right]
\notag\\
&+\left[\frac{g^{2}_{p}}{\delta_{p}}(|\bar{p}\bar{r}\rangle\langle\bar{r}\bar{p}|
+|\bar{r}\bar{p}\rangle\langle\bar{p}\bar{r}|)
+\frac{g^{2}_{q}}{\delta_{q}}(|\bar{q}\bar{r}\rangle\langle\bar{r}\bar{q}|
+|\bar{r}\bar{q}\rangle\langle\bar{q}\bar{r}|)\right]
+\sum_{j=1,2}\Bigg\{\frac{|\Omega^{j}_{0}(t)|^{2}}{\Delta^{j}_{0}}
\left(|\bar{p}\rangle_{jj}\langle\bar{p}|-|\bar{0}\rangle_{jj}\langle\bar{0}|\right)
\notag\\
&+\frac{|\Omega^{j}_{1}(t)|^{2}}{\Delta^{j}_{1}}
\left(|\bar{1}\rangle_{jj}\langle\bar{1}|-|\bar{q}\rangle_{jj}\langle\bar{q}|\right)
+\frac{g^{2}_{p}}{\delta_{p}}\left[\left(1+a^{\dagger}a\right)
|\bar{p}\rangle_{jj}\langle\bar{p}|-a^{\dagger}a|\bar{r}\rangle_{jj}\langle\bar{r}|\right]
+\frac{g^{2}_{q}}{\delta_{q}}\left[\left(1+a^{\dagger}a\right)
|\bar{r}\rangle_{jj}\langle\bar{r}|-a^{\dagger}a|\bar{q}\rangle_{jj}\langle\bar{q}|\right]\Bigg\}, \label{eq1}
\end{align}
\end{widetext}
where the first line of the above equation represents the reduced atom-laser coupling terms, the terms in the square bracket of the second line are the two-atom coupling part, and the terms in the brace of Eq. (\ref{eq1}) are the Stark shifts that can be compensated by introducing ancillary levels \cite{Stark1}.
As shown in Fig. \ref{Fig3},
if the conditions
\begin{align}
&\Delta^{1}_{0}-\delta_{p}=\Delta^{2}_{1}-\delta_{q}>0, \notag\\
&\Delta^{1}_{1}-\delta_{q}=\Delta^{2}_{0}-\delta_{p}>0,
\label{condition}
\end{align}
are satisfied \cite{Ex}, and further the conditions
\begin{align}
&\Delta^{1}_{0}-\delta_{p}\gg \frac{g_{p}|\Omega^{1}_{0}(t)|}{2}\left(\frac{1}{\Delta^{1}_{0}}
+\frac{1}{\delta_{p}}\right),~
\frac{g_{q}|\Omega^{2}_{1}(t)|}{2}\left(\frac{1}{\Delta^{2}_{1}}
+\frac{1}{\delta_{q}}\right),
\notag\\
&\Delta^{1}_{1}-\delta_{q}\gg
\frac{g_{q}|\Omega^{1}_{1}(t)|}{2}\left(\frac{1}{\Delta^{1}_{1}}
+\frac{1}{\delta_{q}}\right),~
\frac{g_{p}|\Omega^{2}_{0}(t)|}{2}\left(\frac{1}{\Delta^{2}_{0}}
+\frac{1}{\delta_{p}}\right), \label{condition1}
\end{align}
are also satisfied, then
the single atom transitions $|\bar{0}\rangle_{j}\leftrightarrow|\bar{r}\rangle_{j}$ and $|\bar{1}\rangle_{j}\leftrightarrow|\bar{r}\rangle_{j}$ can be strongly suppressed while the double atom transitions $|\bar{0}\bar{1}\rangle\leftrightarrow|\bar{r}\bar{r}\rangle$ and $|\bar{1}\bar{0}\rangle\leftrightarrow|\bar{r}\bar{r}\rangle$ are allowed by exchanging virtual photons between two superatoms through a common cavity mode. Thus, the three-level setup coupling is realized, and in this case the effective Hamiltonian reads
\begin{align}
\mathcal{H}_{\mathrm{eff}}(t)=\Omega_{01}(t)|\bar{r}\bar{r}\rangle\langle\bar{0}\bar{1}|
+\Omega_{10}(t)|\bar{r}\bar{r}\rangle\langle\bar{1}\bar{0}|+\mathrm{H.c.}, \label{hamiltonian2}
\end{align}
where
\begin{align}
\Omega_{01}(t)=\frac{g_{p}g_{q}[\Omega^{1}_{0}(t)]^{\ast}\Omega^{2}_{1}(t)}
{4\left(\Delta^{1}_{0}-\delta_{p}\right)}
\left(\frac{1}{\Delta^{1}_{0}}+\frac{1}{\delta_{p}}\right)
\left(\frac{1}{\Delta^{2}_{1}}+\frac{1}{\delta_{q}}\right),
\notag\\
\Omega_{10}(t)=\frac{g_{p}g_{q}[\Omega^{2}_{0}(t)]^{\ast}\Omega^{1}_{1}(t)}
{4\left(\Delta^{1}_{1}-\delta_{q}\right)}
\left(\frac{1}{\Delta^{1}_{1}}+\frac{1}{\delta_{q}}\right)
\left(\frac{1}{\Delta^{2}_{0}}+\frac{1}{\delta_{p}}\right).
\end{align}
Here, the Stark shifts have been removed  by introducing ancillary levels, and the additional two-atom coupling terms in Eq. (\ref{eq1}) have been neglected since they act trivially on the computational space $\mathcal{S}^{\prime}$. Note that the Stark shifts include the terms in Eq. (\ref{eq1}) and the terms generated by the reduced atom laser coupling terms of Eq. (\ref{eq1}) (see \cite{Stark2}).

One remarkable feature of the effective Hamiltonian in Eq. (\ref{hamiltonian2}) is that the double-atom transitions are disentangled from the cavity mode so that our scheme is insensitive to the cavity decay.

To realize the two-qubit nonadiabatic holonomic gate, we set $\Omega_{01}=\Omega^{\prime}(t)\cos(\alpha/2)$ and $\Omega_{10}=\Omega^{\prime}(t)\sin(\alpha/2)$,
where $\Omega^{\prime}(t)$ is time dependent and $\alpha$ is time independent. In this case, the evolution operator generated by the effective Hamiltonian $\mathcal{H}_{\mathrm{eff}}(t)$ can be  written as $U^{\prime}(t)=\exp[-i\int^{t}_{o}\mathcal{H}_{\mathrm{eff}}(t^{\prime})dt^{\prime}]$, and in the basis
$\{|\bar{0}\bar{0}\rangle,|\bar{0}\bar{1}\rangle,
|\bar{1}\bar{0}\rangle,|\bar{1}\bar{1}\rangle,|\bar{r}\bar{r}\rangle\}$, it reads
\begin{widetext}
\begin{align}
U^{\prime}(t)=\left(
  \begin{array}{ccccc}
  1 & 0 & 0 & 0 &0 \\
  0 & \sin^{2}\frac{\alpha}{2}+\cos\gamma_{t}\cos^{2}\frac{\alpha}{2} & \sin\frac{\alpha}{2}\cos\frac{\alpha}{2}(\cos\gamma_{t}-1) & 0 &  -i\sin\gamma_{t}\cos\frac{\alpha}{2}\\
  0 & \sin\frac{\alpha}{2}\cos\frac{\alpha}{2}(\cos\gamma_{t}-1) & \cos^{2}\frac{\alpha}{2}+\cos\gamma_{t}\sin^{2}\frac{\alpha}{2} & 0 &
  -i\sin\gamma_{t}\sin\frac{\alpha}{2}\\
  0 & 0 & 0 & 1 & 0 \\
  0 & -i\sin\gamma_{t}\cos\frac{\alpha}{2} & -i\sin\gamma_{t}\sin\frac{\alpha}{2} & 0 &
  \cos\gamma_{t}\\
  \end{array}
\right)
\end{align}
\end{widetext}
with $\gamma_{t}=\int^{t}_{0}\Omega^{\prime}(t^{\prime})dt^{\prime}$.
If the evolution period $\tau$ satisfies
\begin{align}
\gamma_{\tau}=\int^{\tau}_{0}\Omega^{\prime}(t)dt=\pi, \label{period1}
\end{align}
the evolution operator $U^{\prime}(\tau)$ turns into
\begin{align}
U^{\prime}(\tau)=\left(
  \begin{array}{ccccc}
  1 & 0 & 0 & 0 &0 \\
  0 & -\cos\alpha & -\sin\alpha & 0 &  0 \\
  0 & -\sin\alpha & \cos\alpha & 0 &  0 \\
  0 & 0 & 0 & 1 & 0 \\
  0 & 0 & 0 & 0 & -1 \\
  \end{array}
\right).\label{Tong1}
\end{align}
Equation (\ref{Tong1}) shows that if the initial state is in the computational space $\mathcal{S}^{\prime}$, it will evolve back to $\mathcal{S}^{\prime}$ after the whole evolution, i.e., the condition $(\mathrm{i})$ is satisfied. With the aid of the commutation relation $[\mathcal{H}_{\mathrm{eff}}(t),U^{\prime}(t)]=0$, one can also verify that the parallel transport condition $(\mathrm{ii})$,
\begin{align}
\langle \mu(t)|\mathcal{H}_{\mathrm{eff}}(t)|\nu(t)\rangle=\langle \mu|\mathcal{H}_{\mathrm{eff}}(t)|\nu\rangle=0,~~|\mu\rangle,|\nu\rangle\in\mathcal{S^{\prime}},
\end{align}
is satisfied, where $|\mu(t)\rangle=U^{\prime}(t)|\mu\rangle$ and $|\nu(t)\rangle=U^{\prime}(t)|\nu\rangle$. Therefore, the non-Abelian unitary transformation acting on the computational space $\mathcal{S}^{\prime}$ plays the role of a two-qubit nonadiabatic holonomic gate, and in the basis
$\{|\bar{0}\bar{0}\rangle,|\bar{0}\bar{1}\rangle,|\bar{1}\bar{0}\rangle,|\bar{1}\bar{1}\rangle\}$, it can be written as
\begin{align}
U^{\prime}_{L}(\tau)=\left(
  \begin{array}{cccc}
   1 & 0 & 0 & 0\\
   0 & -\cos\alpha & -\sin\alpha & 0\\
   0 & -\sin\alpha &  \cos\alpha & 0\\
   0 & 0 & 0& 1\\
  \end{array}
\right).
\end{align}
One can readily verify that the above two-qubit gate is nontrivial.

\section{Discussions}

So far, we have realized a universal set of nonadiabatic holonomic gates based on Rydberg superatoms. In the following, we discuss the feasibility of these gates.

As shown in Ref. \cite{Saffman2005}, the lifetime $\tau_{c}$ of the Rydberg state of the $\mathrm{Rb}$ atom with principal quantum number $n=95$ can be up to $300\mathrm{\mu s}$ and the dipole-dipole interaction strength between atoms is above $2\pi\times30\mathrm{GHz}$ when atoms are separated less than $2\mathrm{\mu m}$. Suppose each of our Rydberg superatoms consists of $N=100$ identical atoms within blockade radius $2\mathrm{\mu m}$.

\begin{figure}[t]
  \includegraphics[scale=0.45]{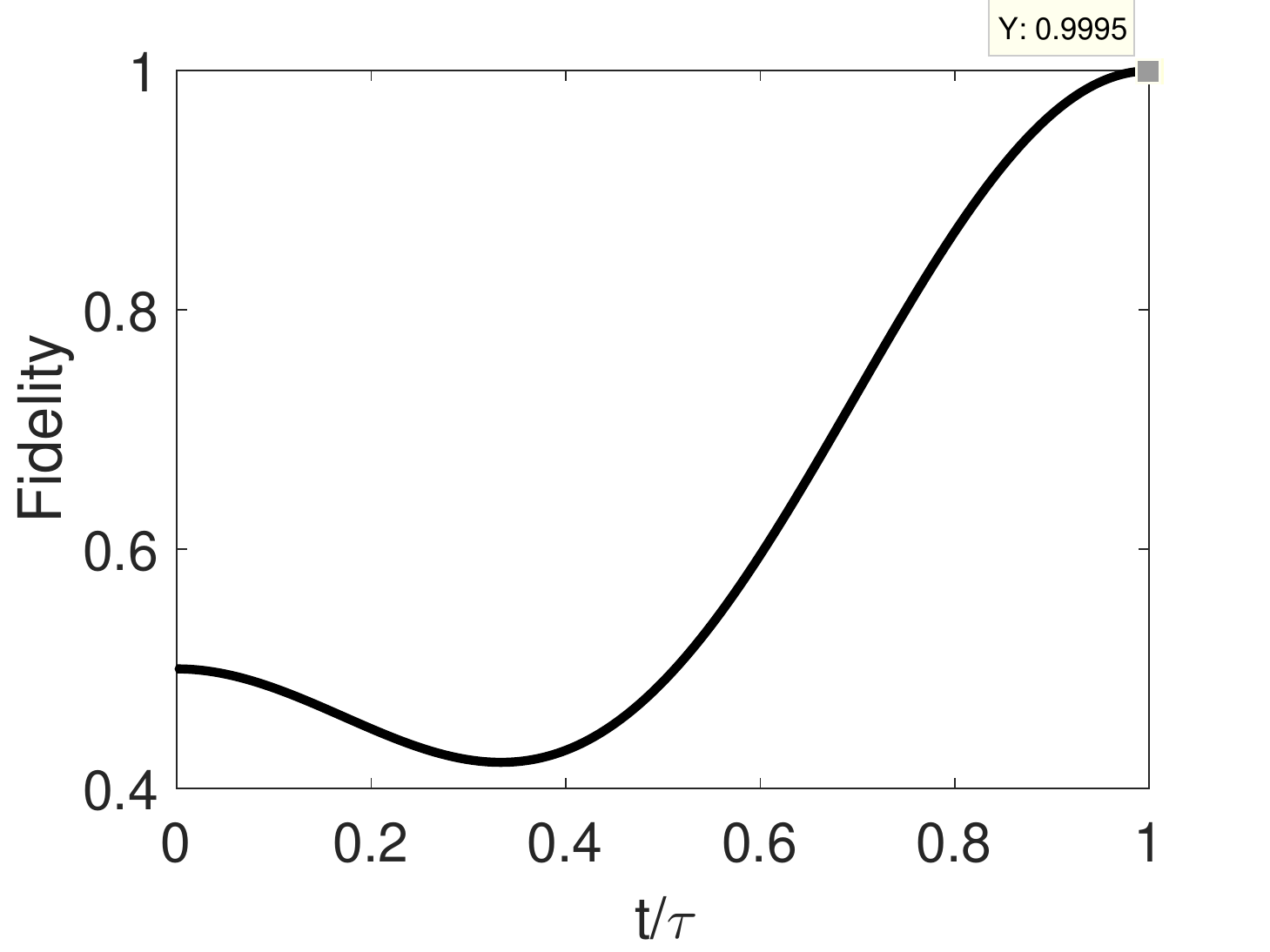}
  \caption{Fidelity dynamics as a function of $t/\tau$ for the gate $U(\tau)=\exp(-i\pi\sigma_{x}/4)$ with initial state $|\bar{0}\rangle$. The fidelity is up to $99.95\%$ under the influence of decay. Here, the parameters of the laser pulses are taken as $\Delta=-2\pi\times5\mathrm{MHz}$,
$\Omega_{0}(t)=\Omega_{1}(t)=2\pi\times1.25\sqrt{3}\mathrm{MHz}$, and the decay ratio is taken as $\gamma=2\pi\times(1/\tau_{c})\mathrm{MHz}$.} \label{Fig4}
\end{figure}

For the one-qubit gates, the evolution time is determined by the detuning $\Delta$ and the Rabi frequencies $\Omega_{0}(t)$ and $\Omega_{1}(t)$. It can be shown that by properly choosing $\Delta$, $\Omega_{0}(t)$, and $\Omega_{1}(t)$, the evolution time $\tau$ can be much shorter than the lifetime $\tau_{c}$. For example, the detuning and the Rabi frequencies can be chosen as
$\Delta\sim2\pi\times10\mathrm{MHz}\cdot\sin\theta$,
$\Omega_{0}(t)\sim2\pi\times5\mathrm{MHz}\cdot\cos\theta\cos(\varphi/2)$, and $\Omega_{1}(t)\sim 2\pi\times5\mathrm{MHz}\cdot\cos\theta\sin(\varphi/2)$, where $\theta\in[0,\pi]$ and $\varphi\in[0,2\pi]$. The above parameters, being much smaller than the dipole-dipole interaction strength, satisfy the requirement of Rydberg blockade and are experimentally allowed. By using these parameters, we can estimate the evolution time of the one-qubit gates, and it is $\tau\sim 0.1\mathrm{\mu s}$, which is sufficiently short compared with the lifetime, $\tau_{c}\sim300\mathrm{\mu s}$.
Furthermore, our numerical result indicates that the fidelity is up to $99.95\%$ under the influence of decay for the gate $U(\tau)=\exp(-i\pi\sigma_{x}/4)$ if the initial state is taken as $|\bar{0}\rangle$, shown in Fig. \ref{Fig4}.

\begin{figure*}[t]
  \includegraphics[scale=0.4]{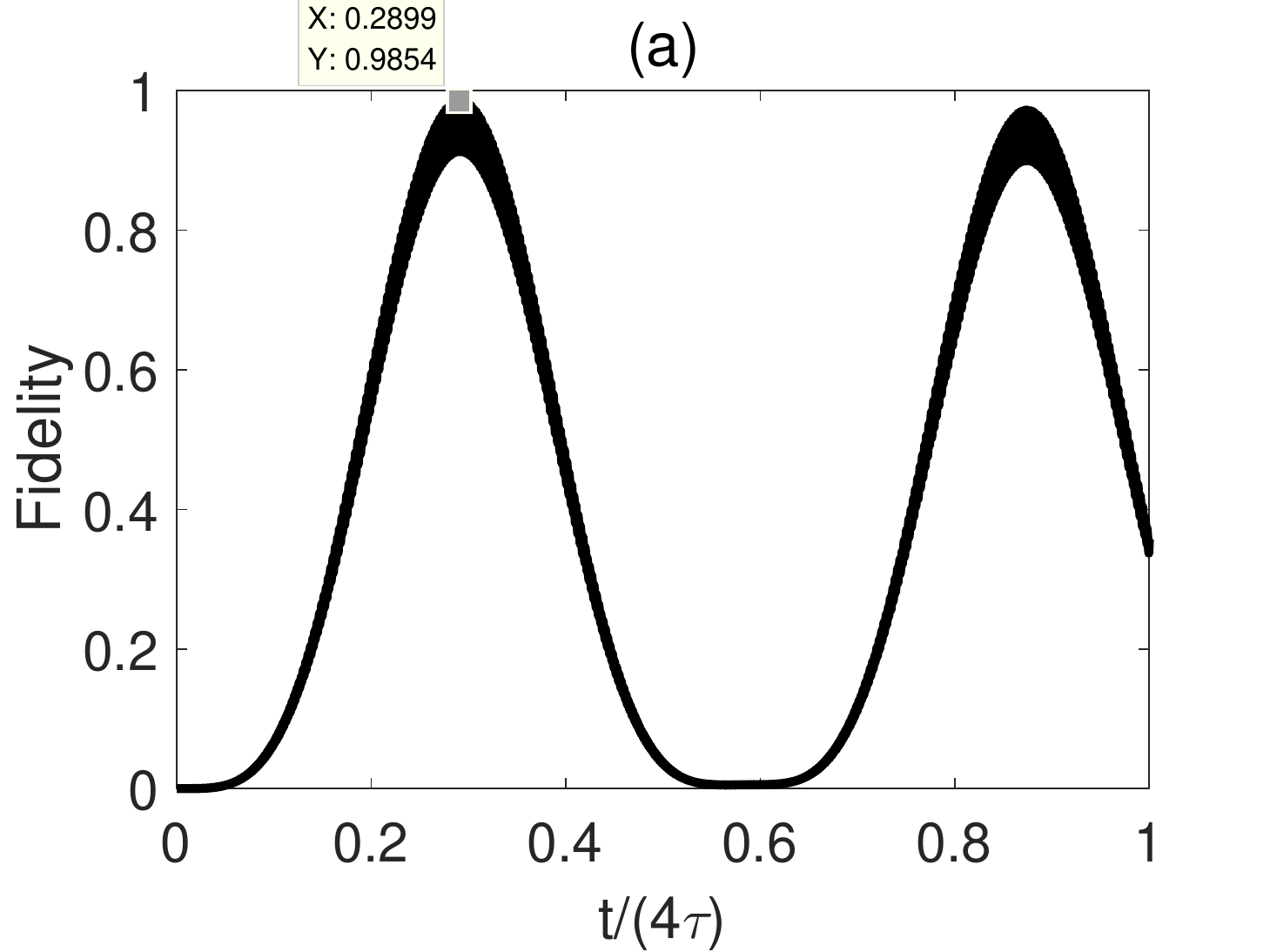}
  \includegraphics[scale=0.4]{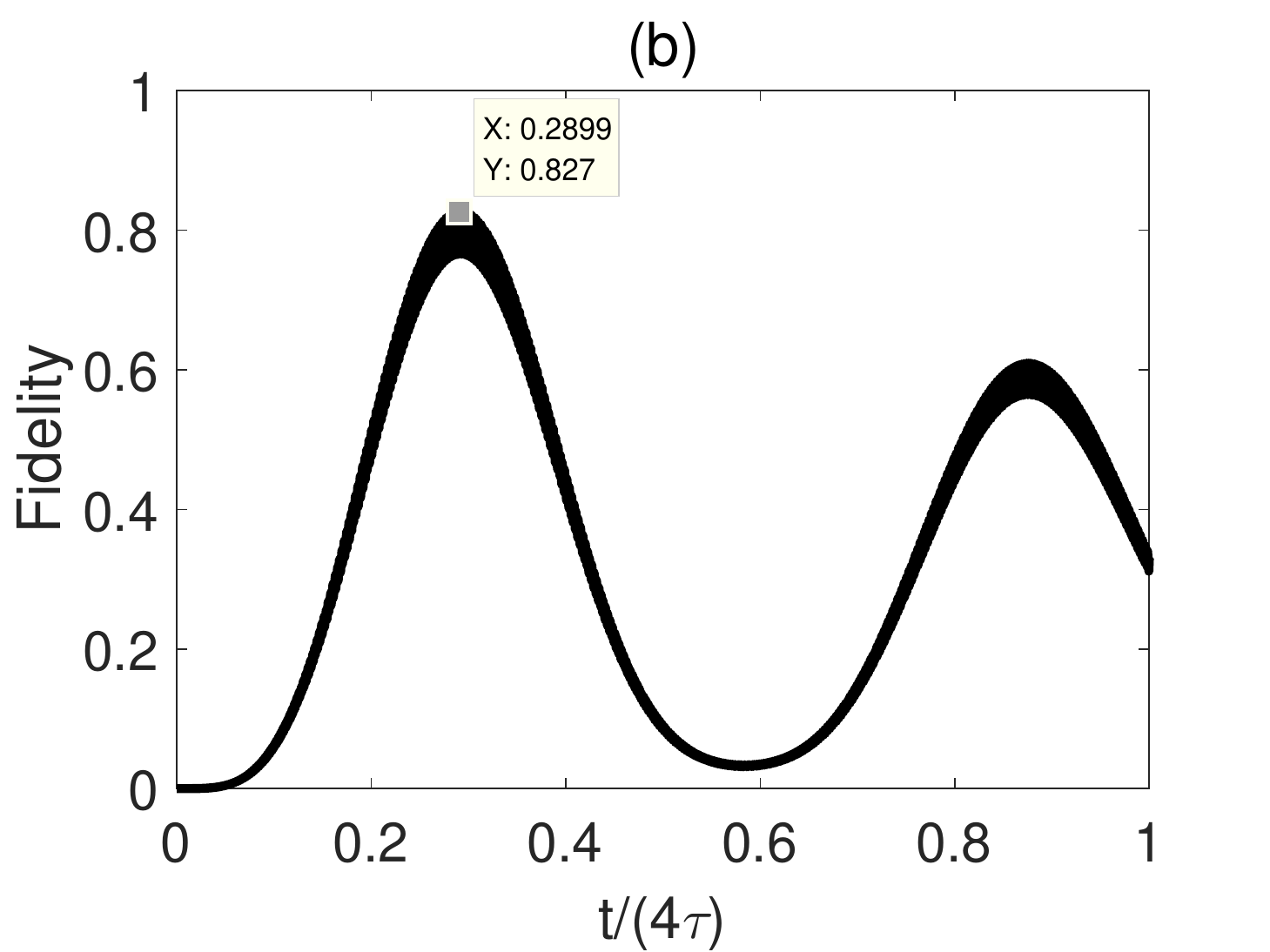}
  \includegraphics[scale=0.4]{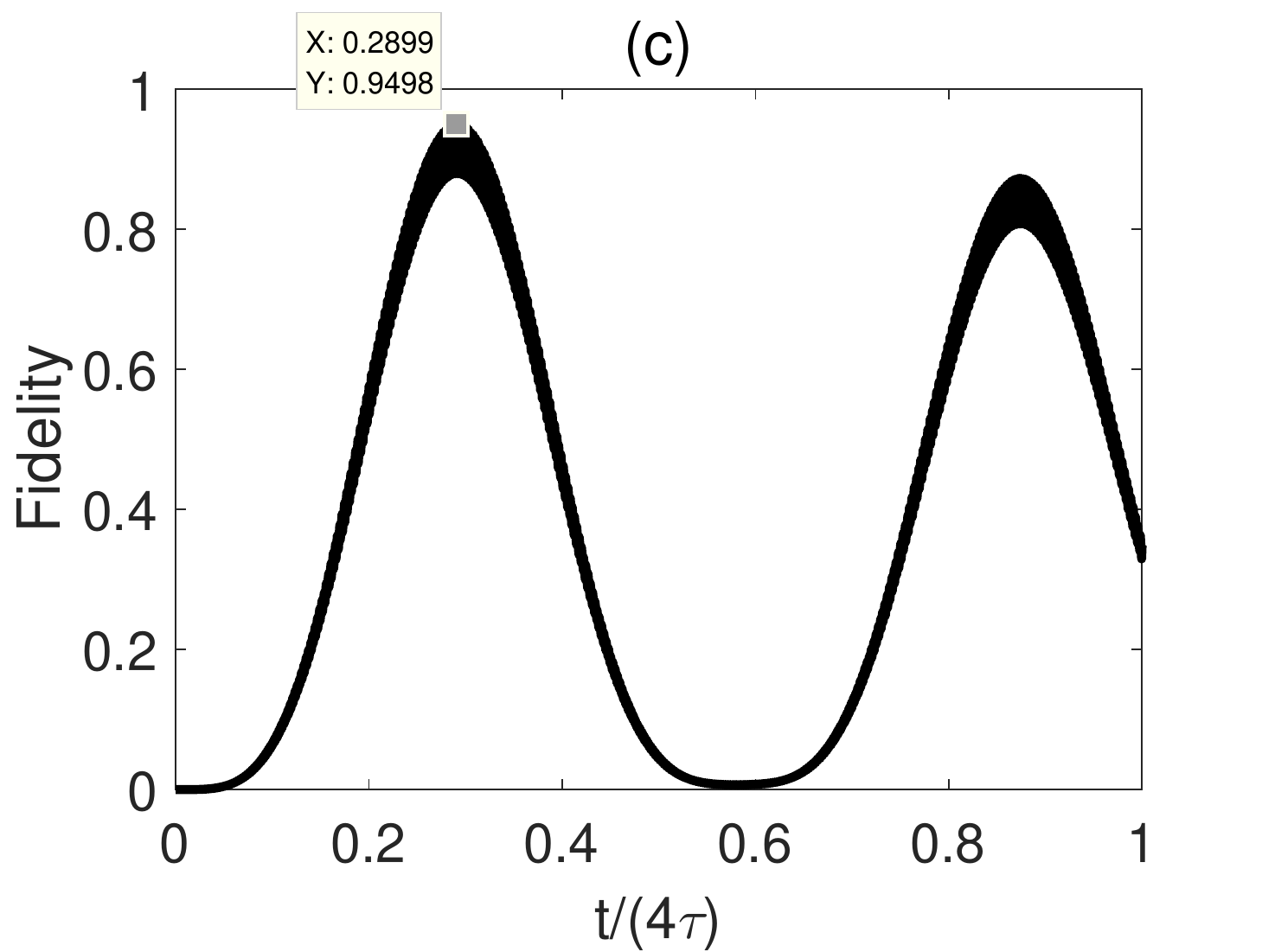}
  \caption{Fidelity dynamics as a function of $t/(4\tau)$ for the gate $U^{\prime}_{L}(\tau)=|\bar{0}\bar{0}\rangle\langle\bar{0}\bar{0}|-|\bar{0}\bar{1}\rangle\langle\bar{1}\bar{0}|
  -|\bar{1}\bar{0}\rangle\langle\bar{0}\bar{1}|+|\bar{1}\bar{1}\rangle\langle\bar{1}\bar{1}|$ with initial state $|\bar{0}\bar{1}\rangle$. (a) The performance of the real Hamiltonian in Eq. {(\ref{hamiltonian})}.
  The fidelity between the result obtained by the real Hamiltonian and the desired theoretical result is up to $98.54\%$.
  (b) The performance of the gate under the influence of decay. The result shows that the fidelity is $82.70\%$, where the decay ratio is taken as $\gamma=2\pi\times(1/\tau_{c})\mathrm{MHz}$ and the decay operators are taken as $L_{0p}=\sum^{2}_{j=1}|\bar{0}\rangle_{jj}\langle \bar{p}|$, $L_{1p}=\sum^{2}_{j=1}|\bar{1}\rangle_{jj}\langle \bar{p}|$, $L_{0q}=\sum^{2}_{j=1}|\bar{0}\rangle_{jj}\langle \bar{q}|$, $L_{1q}=\sum^{2}_{j=1}|\bar{1}\rangle_{jj}\langle \bar{q}|$, $L_{0r}=\sum^{2}_{j=1}|\bar{0}\rangle_{jj}\langle \bar{r}|$, and $L_{1r}=\sum^{2}_{j=1}|\bar{1}\rangle_{jj}\langle \bar{r}|$.
  (c) The performance of the gate under the influence of decay when the parameters of laser
   pulses as well as the cavity field are increased to five times the early values. The numerical result shows that the fidelity is up to $94.98\%$.} \label{Fig5}
\end{figure*}

For the two-qubit gate, the evolution time of the gates is determined by the parameter $\Omega^{\prime}(t)$, which is further restricted by the parameters of the laser pulses and cavity field (see Sec. III). By properly choosing the parameters of the laser pulses and the cavity field, one can find that the evolution time $\tau\ll\tau_{c}$. For example, the Rabi frequencies and the coupling constants can be chosen as $\Omega^{1}_{0}(t)=\Omega^{2}_{1}(t)=g_{p}=g_{q}=2\pi\times10\mathrm{MHz}$, $\Omega^{1}_{1}(t)= 2\pi\times14.5\mathrm{MHz}$, and $\Omega^{2}_{0}(t)=2\pi\times14.6888\mathrm{MHz}$. The detunings can be chosen as $\delta_{p}=2\delta_{q}=2\pi\times200\mathrm{MHz}$, $\bar{\Delta}^{1}_{0}=2\pi\times210\mathrm{MHz}$,
$\bar{\Delta}^{2}_{0}=2\pi\times220\mathrm{MHz}$, $\bar{\Delta}^{1}_{1}=2\pi\times120\mathrm{MHz}$,
and $\bar{\Delta}^{2}_{1}=2\pi\times110\pi\mathrm{MHz}$. By using the above parameters, one can get $\Omega^{\prime}(t)=2\pi\times0.0659\mathrm{MHz}$ and then the evolution time of our gate is $\tau\sim 7.59\mathrm{\mu s}$, which is sufficiently short compared with the lifetime, $\tau_{c}\sim300\mathrm{\mu s}$. It is worth noting that the chosen parameters satisfy the requirement of the Rydberg blockade as well as other requirements for realizing the effective Hamiltonian in Eq. (\ref{hamiltonian2}), and are experimentally available. By using numerical simulation, we demonstrate the performance of the real Hamiltonian in Eq. {(\ref{hamiltonian})}. The fidelity between the result obtained by the real Hamiltonian and the desired theoretical result can be up to $98.54\%$ for the gate $U^{\prime}_{L}(\tau)=|\bar{0}\bar{0}\rangle\langle\bar{0}\bar{0}|
-|\bar{0}\bar{1}\rangle\langle\bar{1}\bar{0}|-|\bar{1}\bar{0}\rangle\langle\bar{0}\bar{1}|
+|\bar{1}\bar{1}\rangle\langle\bar{1}\bar{1}|$, where the initial state is taken as $|\bar{0}\bar{1}\rangle$, shown in Fig. \ref{Fig5}(a). Subsequently, we examine the performance of the gate under the influence of decay. Our result indicates that the fidelity is $82.70\%$, shown in Fig \ref{Fig5}(b). It means that the decay seriously influences our gate for the present parameters. To improve the performance of the gate under the influence of decay, we increase the parameters of laser pulses as well as the cavity field to five times the above values. The numerical simulation indicates that the fidelity is up to $94.98\%$, shown in Fig. \ref{Fig5}(c).

In general, the larger the ratio of detunings to coupling strengths is, the better the Hamiltonian in Eq. (\ref{hamiltonian}) performs. Yet, the larger ratio leads to lower effective Rabi frequencies and further leads to slower gates. It ultimately results in a stronger influence of decoherence on gates. Our numerical simulation shows that the present ratio of detunings to coupling strengths leads to good performance of the Hamiltonian in Eq. (\ref{hamiltonian}) but low effective Rabi frequencies and slow gates, which results in a serious influence of decay on gates. Here, the lifetime of Rydberg states is 40 times the evolution time of the gate. To reduce the influence of the decay on the gates, the parameters of laser pulses as well as the cavity field are increased to five times the early values. In this case, the effective Rabi frequencies are increased to five times the early Rabi frequencies, and the speed of gates is also increased to five times the early speed, which makes the influence of the decay greatly reduced. Here, the lifetime of Rydberg states is 200 times the evolution time of the gate.

\section{Conclusion}

In conclusion, we have proposed a scheme of nonadiabatic holonomic quantum computation with mesoscopic atomic ensembles. By encoding a qubit into a pair of collective ground states of a Rydberg superatom, we realize a universal set of nonadiabatic holonomic gates. The one-qubit gates are performed by off-resonant laser pulses. The two-qubit gate is performed with the aid of a microwave cavity. The transitions between the double collective ground states of Rydberg superatoms and the double collective Rydberg states are facilitated by exchanging virtual photons through a common cavity mode. In this process, the effective double atom transitions are disentangled from the cavity mode and therefore the two-qubit gate is insensitive to the cavity decay. It is interesting to note that besides the common merits of nonadiabatic holonomic quantum computation such as the robustness and the speediness, our Rydberg-superatom-based scheme has two particular merits: the long coherence time of Rydberg states and the operability of the mesoscopic systems.

\begin{acknowledgments}
P.Z.Z. thanks T. Chen, S. L. Su, and J. L. Wu for helpful discussions.
P.Z.Z. acknowledges support from the National Natural Science Foundation of China through Grant No. 11575101. X.W. and T.H.X. acknowledge support from the National Natural Science Foundation of China through Grant No. 11775129. G.F.X. acknowledges support from the National Natural Science Foundation of China through Grant No. 11605104, and from the Future Project for Young Scholars of Shandong University through Grant No. 2016WLJH21. D.M.T. acknowledges support from the National Basic Research Program of China through Grant No. 2015CB921004.
\end{acknowledgments}


\begin{thebibliography}{99}
\bibitem{Sjoqvist2012} E. Sj\"{o}qvist, D. M. Tong, L. M. Andersson, B. Hessmo, M. Johansson, and K. Singh, New J. Phys. \textbf{14}, 103035 (2012).
\bibitem{Xu2012} G. F. Xu, J. Zhang, D. M. Tong, E. Sj\"{o}qvist, and L. C. Kwek, Phys. Rev. Lett. \textbf{109}, 170501 (2012).
\bibitem{Anandan} J. Anandan, Phys. Lett. A \textbf{133}, 171 (1988).
\bibitem{Wilczek} F. Wilczek and A. Zee, Phys. Rev. Lett. \textbf{52}, 2111 (1984).
\bibitem{Zanardi} P. Zanardi and M. Rasetti, Phys. Lett. A \textbf{264}, 94 (1999).
\bibitem{Duan} L. M. Duan, J. I. Cirac, and P. Zoller, Science \textbf{292}, 1695 (2001).
\bibitem{Johansson2012} M. Johansson, E. Sj\"{o}qvist, L. M. Andersson, M. Ericsson, B. Hessmo, K. Singh, and D. M. Tong, Phys. Rev. A \textbf{86}, 062322 (2012).
\bibitem{Spiegelberg2013} J. Spiegelberg and E. Sj\"{o}qvist, Phys. Rev. A \textbf{88}, 054301 (2013).
\bibitem{Zhang2014} J. Zhang, L. C. Kwek, E. Sj\"{o}qvist, D. M. Tong, and P. Zanardi, Phys. Rev. A \textbf{89}, 042302 (2014).
\bibitem{Mousolou2014} V. A. Mousolou, C. M. Canali, and E. Sj\"{o}qvist, New J. Phys. \textbf{16}, 013029 (2014).
\bibitem{Xu2015} G. F. Xu, C. L. Liu, P. Z. Zhao, and D. M. Tong, Phys. Rev. A \textbf{92}, 052302 (2015).
\bibitem{Sjovist2016} E. Sj\"{o}qvist, Phys. Lett. A \textbf{380}, 65 (2016).
\bibitem{Sjovist2016PRA} E. Herterich and E. Sj\"{o}qvist, Phys. Rev. A \textbf{94}, 052310 (2016).
\bibitem{Sun2016} C. F. Sun, G. C. Wang, C. F. Wu, H. D. Liu, X. L. Feng, J. L. Chen, and K. Xue, Sci. Rep. \textbf{6}, 20292 (2016).
\bibitem{Liang2014} Z. T. Liang, Y. X. Du, W. Huang, Z. Y. Xue, and H. Yan, Phys. Rev. A \textbf{89}, 062312 (2014).
\bibitem{Zhou2015}  J. Zhou, W. C. Yu, Y. M. Gao, and Z. Y. Xue, Opt. Express \textbf{23}, 14027 (2015).
\bibitem{Xue2015} Z. Y. Xue, J. Zhou, and Z. D. Wang, Phys. Rev. A \textbf{92}, 022320 (2015).
\bibitem{You2016} Y. M. Wang, J. Zhang, C. F. Wu, J. Q. You, and G. Romero, Phys. Rev. A \textbf{94}, 012328 (2016).
\bibitem{Xue2016} Z. Y. Xue, J. Zhou, Y. M. Chu, and Y. Hu, Phys. Rev. A \textbf{94}, 022331 (2016).
\bibitem{Xue2017} Z. Y. Xue, F. L. Gu, Z. P. Hong, Z. H. Yang, D. W. Zhang, Y. Hu, and J. Q. You, Phys. Rev. Appl. \textbf{7}, 054022 (2017).
\bibitem{Xue2017PRA} B. J. Liu, Z. H. Huang, Z. Y. Xue, and X. D. Zhang, Phys. Rev. A \textbf{95}, 062308 (2017).
\bibitem{Zhao} P. Z. Zhao, X. D. Cui, G. F. Xu, E. Sj\"{o}qvist, and D. M. Tong, Phys. Rev. A \textbf{96}, 052316 (2017).
\bibitem{Xia2018} Y. H. Kang, Y. H. Chen, Z. C. Shi, B. H. Huang, J. Song, and Y. Xia, Phys. Rev. A \textbf{97}, 042336 (2018).
\bibitem{Zhao2017} P. Z. Zhao, G. F. Xu, Q. M. Ding, E. Sj\"{o}qvist, and D. M. Tong, Phys. Rev. A \textbf{95}, 062310 (2017).
\bibitem{Xu2017} G. F. Xu, P. Z. Zhao, T. H. Xing, E. Sj\"{o}qvist, and D. M. Tong, Phys. Rev. A \textbf{95}, 032311 (2017).
\bibitem{Xu2017PRA} G. F. Xu, P. Z. Zhao, D. M. Tong, and E. Sj\"{o}qvist, Phys. Rev. A \textbf{95}, 052349 (2017).
\bibitem{Mousolou2017} V. A. Mousolou, Phys. Rev. A \textbf{96}, 012307 (2017).
\bibitem{Long} G. R. Feng, G. F. Xu, and G. L. Long, Phys. Rev. Lett. \textbf{110}, 190501 (2013).
\bibitem{Abdumalikov} A. A. Abdumalikov, J. M. Fink, K. Juliusson, M. Pechal, S. Berger, A. Wallraff, and S. Filipp, Nature(London) \textbf{496}, 482 (2013).
\bibitem{Arroyo} S. Arroyo-Camejo, A. Lazariev, S. W. Hell, and G. Balasubramanian, Nat. Commun. \textbf{5}, 4870 (2014).
\bibitem{Duan2014} C. Zu, W. B. Wang, L. He, W. G. Zhang, C. Y. Dai, F. Wang, and L. M. Duan, Nature(London) \textbf{514}, 72 (2014).
\bibitem{Zhou2017} Brian B. Zhou, Paul C. Jerger, V. O. Shkolnikov, F. J. Heremans, Guido Burkard, and David D. Awschalom, Phys. Rev. Lett. \textbf{119}, 140503 (2017).
\bibitem{Long2017} H. Li, Y. Liu, and G. L. Long, Sci. China-Phys. Mech. Astron. \textbf{60}, 080311 (2017).
\bibitem{Lukin2001} M. D. Lukin, M. Fleischhauer, R. Cote, L. M. Duan, D. Jaksch, J. I. Cirac, and P. Zoller, Phys. Rev. Lett. \textbf{87}, 037901 (2001).
\bibitem{Tong2004} D. Tong, S. M. Farooqi, J. Stanojevic, S. Krishnan, Y. P. Zhang, R. C\^{o}t\'{e}, E. E. Eyler, and P. L. Gould, Phys. Rev. Lett. \textbf{93}, 063001 (2004).
\bibitem{Heidemann2007} R. Heidemann, U. Raitzsch, V. Bendkowsky, B. Butscher, R. L\"{o}w, L. Santos, and T. Pfau, Phys. Rev. Lett. \textbf{99}, 163601 (2007).
\bibitem{Bakr2009} Waseem S. Bakr, Jonathon I. Gillen, A. Peng, S. F\"{o}lling, and M. Greiner, Nature (London) \textbf{462}, 74 (2009).
\bibitem{Ebert2015} M. Ebert, M. Kwon, T. G. Walker, and M. Saffman, Phys. Rev. Lett. \textbf{115}, 093601 (2015).
\bibitem{Dur2000} W. D\"{u}r, G. Vidal, and J. I. Cirac, Phys. Rev. A \textbf{62}, 062314 (2000).
\bibitem{Brion2007} E. Brion, K. M{\o}lmer, and M. Saffman, Phys. Rev. Lett. \textbf{99}, 260501 (2007).
\bibitem{Muller2009} M. M\"{u}ller, I. Lesanovsky, H. Weimer, H. P. B\"{u}chler, and P. Zoller, Phys. Rev. Lett. \textbf{102}, 170502 (2009).
\bibitem{Han2010} Y. Han, B. He, K. Heshami, Cheng-Zu Li, and C. Simon, Phys. Rev. A \textbf{81}, 052311 (2010).
\bibitem{Wu2010} H. Z. Wu, Z. B. Yang, and S. B. Zheng, Phys. Rev. A \textbf{82}, 034307 (2010).
\bibitem{Dudin2012} Y. O. Dudin and A. Kuzmich, Science \textbf{336}, 887 (2012).
\bibitem{Beterov2013} I. I. Beterov, M. Saffman, E. A. Yakshina, V. P. Zhukov, D. B. Tretyakov, V. M. Entin, I. I. Ryabtsev, C. W. Mansell, C. MacCormick, S. Bergamini, and M. P. Fedoruk, Phys. Rev. A \textbf{88}, 010303(R) (2013).
\bibitem{Weber2015} T. M. Weber, M. H\"{o}ning, T. Niederpr\"{u}m, T. Manthey, O. Thomas, V. Guarrera, M. Fleischhauer, G. Barontini, and H. Ott, Nat. Phys. \textbf{11}, 157 (2015).
\bibitem{Zeiher2015} J. Zeiher, P. Schau{\ss}, S. Hild, T. Macr\`{i}, I. Bloch, and C. Gross, Phys. Rev. X \textbf{5}, 031015 (2015).
\bibitem{Sarkany2015} L. S\'{a}rk\'{a}ny, J. Fort\'{a}gh, and D. Petrosyan, Phys. Rev. A \textbf{92}, 030303(R) (2015).
\bibitem{Jaksch2000} D. Jaksch, J. I. Cirac, P. Zoller, S. L. Rolston, R. C\^{o}t\'{e}, and M. D. Lukin, Phys. Rev. Lett. \textbf{85}, 2208 (2000).
\bibitem{James2007} D. F. V. James and J. Jerke,  Can. J. Phys. \textbf{85}, 625 (2007).
\bibitem{Stark1} In the case of the initial state of the vibrational mode being a vacuum state, the Stark shifts $\sum_{j=1,2}\Big[\frac{g^{2}_{p}}{\delta_{p}}(|\bar{p}\rangle_{jj}\langle \bar{p}|-|\bar{r}\rangle_{jj}\langle \bar{r}|)+\frac{g^{2}_{q}}{\delta_{q}}(|\bar{r}\rangle_{jj}\langle \bar{r}|-|\bar{q}\rangle_{jj}\langle \bar{q}|)\Big]a^{\dagger}a$ are equal to zero so that there is no need to compensate these terms.
For the Stark shifts $\sum_{j=1,2}\bigg\{\Big[\frac{|\Omega^{j}_{0}(t)|^{2}}{\Delta^{j}_{0}}
+\frac{g^{2}_{p}}{\delta_{p}}\Big]|\bar{p}\rangle_{jj}\langle \bar{p}|-\frac{|\Omega^{j}_{1}(t)|^{2}}{\Delta^{j}_{1}}|\bar{q}\rangle_{jj}\langle \bar{q}|\bigg\}$, it is not necessary to  compensate them because the states $|\bar{p}\rangle_{j}$ and $|\bar{q}\rangle_{j}$ are decoupled with the basis $\{|\bar{0}\rangle_{j},|\bar{1}\rangle_{j}\}$ for the reduced Hamiltonian in Eq. (\ref{eq1}) and thus the Stark shifts cannot influence the atom-laser coupling terms of the reduced Hamiltonian.
For the Stark shifts $\sum_{j=1,2}\Big[-\frac{|\Omega^{j}_{0}(t)|^{2}}{\Delta^{j}_{0}}|\bar{0}\rangle_{jj}\langle\bar{0}|
+\frac{|\Omega^{j}_{1}(t)|^{2}}{\Delta^{j}_{1}}|\bar{1}\rangle_{jj}\langle\bar{1}|\Big]$, we can introduce laser pulses to generate the couplings $\sum_{j=1,2}\Big[\Omega^{j}_{0}(t)e^{-i\Delta^{j}_{0}t}
|\bar{p}\rangle_{jj}\langle\bar{0}|
+\Omega_{1}^{j}(t)e^{i\Delta^{j}_{1}t}|\bar{q}\rangle_{jj}\langle\bar{1}|+\mathrm{H.c.}\Big]$, which leads to the opposite Stark shifts $\sum_{j=1,2}\Big[\frac{|\Omega^{j}_{0}(t)|^{2}}{\Delta^{j}_{0}}|\bar{0}\rangle_{jj}\langle\bar{0}|
-\frac{|\Omega^{j}_{1}(t)|^{2}}{\Delta^{j}_{1}}|\bar{1}\rangle_{jj}\langle\bar{1}|\Big]$ in the condition of large detuning approximations and thus can compensate the unwanted Stark shifts.
For the rest of the Stark shifts $\sum_{j=1,2}\Big(\frac{g^{2}_{q}}{\delta_{q}}|\bar{r}\rangle_{jj}\langle \bar{r}|\Big)$, we can introduce ancillary levels $|\bar{a}\rangle_{j}$ and laser pulses to generate the couplings $\sum_{j=1,2}\Big(g_{q}e^{i\delta_{q}t}|\bar{a}\rangle_{jj}\langle \bar{r}|+\mathrm{H.c.}\Big)$, which leads to the Stark shifts $\sum_{j=1,2}\frac{g^{2}_{q}}{\delta_{q}}(|\bar{a}\rangle_{jj}\langle \bar{a}|-|\bar{r}\rangle_{jj}\langle \bar{r}|)$ and thus can compensate $\sum_{j=1,2}\Big(\frac{g^{2}_{q}}{\delta_{q}}|\bar{r}\rangle_{jj}\langle \bar{r}|\Big)$.
\bibitem{Ex} The cavity detunings $\delta_p$ and $\delta_q$ are treated as two independent parameters, but in fact they are determined by a single parameter, the cavity vibrational frequency. Three adjacent collective Rydberg states $|\bar{p}\rangle$, $|\bar{q}\rangle$, and $|\bar{r}\rangle$ will have a fixed level spacing governed by the principal quantum number $n$, and this will also fix $|\delta_q - \delta_p|$.
\bibitem{Stark2} The additional Stark shifts generated by the atom-laser coupling terms of Eq. (\ref{eq1}) are $\sum_{j}\bigg\{\frac{1}{\Delta^{j}_{0}-\delta_{p}}
\Big[\frac{g_{p}|\Omega^{j}_{0}(t)|}{2}\Big(\frac{1}{\Delta^{j}_{0}}+\frac{1}{\delta_{p}}\Big)\Big]^{2}
\Big[a^{\dagger}a|\bar{r}\rangle_{jj}\langle \bar{r}|-(1+a^{\dagger}a)|\bar{0}\rangle_{jj}\langle\bar{0}|\Big]+
\frac{1}{\Delta^{j}_{1}-\delta_{q}}\Big[\frac{g_{q}|\Omega^{j}_{1}(t)|}{2}\Big(\frac{1}{\Delta^{j}_{1}}
+\frac{1}{\delta_{q}}\Big)\Big]^{2}
\Big[a^{\dagger}a|\bar{1}\rangle_{jj}\langle\bar{1}|-(1+a^{\dagger}a)|\bar{r}\rangle_{jj}\langle \bar{r}|\Big]\bigg\}$.
\bibitem{Saffman2005} M. Saffman and T. G. Walker, Phys. Rev. A \textbf{72}, 022347 (2005).

\end{thebibliography}
\end{document}